\documentstyle[12pt]{article}

\input{tcilatex}

\QQQ{Language}{
American English
}

\begin{document}

\bigskip

\begin{center}
{\huge {\bf Future Destiny of Quintessential Universe and Constraint on
Model from Deceleration Parameter}}

\bigskip \bigskip

{\Large {\bf De-Hai Zhang}}\\[0pt]

\bigskip 

(e-mail: zhang.dh@263.net)\\[0pt]

Department of Physics, Graduate School in Beijing,\\[0pt]

University of Science and Technology of China,\\[0pt]

P.O.Box 3908, Beijing 100039, P.R.China.\\[0pt]

\bigskip

{\bf Abstract:}
\end{center}

{\hspace*{5mm}The evolution of the quintessence in various stages of the
universe, the radiation-, matter-, and quintessence-dominated, is closely
related with the tracking behavior and the deceleration parameter of the
universe. We gave the explicit relation between the equation-of-state of the
quintessence in the epoch of the matter-quintessence equality and the
inverse power index of the quintessence potential, obtained the constraint
on this potential parameter come from the present deceleration parameter,
i.e., a low inverse power index. We point out that the low inverse power-law
potential with a single term can not work for the tracking solution. In
order to have both of the tracker and the suitable deceleration parameter it
is necessary to introduce at least two terms in the quintessence potential.
We give the future evolution of the quintessential universe.}

\bigskip

\section{Introduction}

{\hspace*{5mm}}The important observation on the high redshift Type Ia
supernovas in 1998 is to discover that the universe is expanding accelerated
with the negative deceleration parameter$^{[1,2]}$%
$$
q_0\equiv -\ddot R/(RH^2)=-0.33\pm 0.17.\eqno{(1)} 
$$
The universe must contain some new dark energy which equation-of-state $%
w=p/\rho $ is negative. In spite of the small non-zero cosmology constant ($%
w_\lambda =-1$) is one of the possible explanations, people does not satisfy
this scheme due to its bad coincidence and fine-turning problems. A popular
candidate is the quintessence$^{[3]}$ ($-1<w_q<0$), which is a slowly varied
scalar field $\phi $ and has an inverse power-law potential $V=m^{4+\beta
}\phi ^{-\beta }$, where we call the parameter $\beta $ as the inverse power
index of the quintessence. We see that its minimum, i.e., vacuum has zero
energy, this tallies with such a thought that the true cosmological constant
should be zero due to some unknown profound reason. Zlatev et al. find out
that the quintessence has a nice property, the tracking behavior, and its
potential initial values with almost 100 different orders can converge to a
common evolving final state, i.e., the tracking solution$^{[4]}$. This
scheme can well solve the coincidence problem. The idea of the quintessence
received enough attentions, many papers did further researches on the
quintessence$^{[5,6]}$, and developed many new ideas$^{[7,8]}$. The inverse
power index $\beta $ is an important parameter for quintessence, the
quintessence will not be able to come into the tracking situation up to the
present time if $\beta $ is too small, for example $\beta <5$ as shown by
Ref.[4]. An important problem is whether the present experiment, more
concretely from the cosmic deceleration parameter, have added some
restriction on this inverse power index. In this paper we shall research
this interesting problem in detail. We shall find out the
equations-of-states of the quintessence in various evolving stages of the
universe at first, then express the deceleration parameter in these
equation-of-states. The observation result of the deceleration parameter
will constrain the inverse power index of the quintessence up to $\beta \leq
2$. We shall give the evolution of the quintessence energy  density, the
cosmic scale factor and the equation-of-state of the quintessence in the
future. 

\section{Equation-of-states of quintessence}

{\hspace*{5mm}}The evolution of the quintessence in the radiation dominated
or the matter dominated epochs is dependence on its equation-of-states $%
w_q=(\beta w_B-2)/(\beta +2)$, where $w_B$ is the equation-of-state of the
background. We cite this formula directly, which has obtained in Ref.[4].
The best tracking behavior requires that the evolution velocity of the
quintessence in the radiation dominated epoch ($w_r=1/3$) is the equal to
(or larger than) the one of the matter ($w_m=0$), i.e., $w_q^{(r)}=(\beta
/3-2)/(\beta +2)=0$, then we obtain $\beta =6$. The equation-of-state of the
quintessence in the matter dominated epoch is $w_q^{(m)}=-(\beta /2+1)^{-1}$%
, which evolution is slower than one of the matter.

As the universe expanses, the quintessential potential energy will overweigh
one of the matter and become quintessence dominated, it is important to know
what is the equation-of-state of the quintessence in the quintessence
dominated epoch, which can be obtained by analyzing the equation of field
motion $3H\dot \phi =\beta m^{4+\beta }\phi ^{-\beta -1}$ and $%
3M_p^2H^2=m^{4+\beta }\phi ^{-\beta }$, where $M_p\equiv (8\pi G)^{-1/2}$ $%
=2.4\times 10^{18}$GeV is the Planck energy. Note that in this stage the
field rolls slowly and the potential energy is dominated, these allow us to
take the above approximation. We suppose that the solution is a simple
power-law type of the time, $\phi \propto t^\alpha $. Comparing the
exponential about time $t$ in the two sides of the equations, we get $\alpha
=2/(\beta +4)$. This power value confirms further the conditions of the slow
rolling $\ddot \phi \ll 3H\dot \phi $ and potential dominated $\dot \phi
^2/2\ll V$, and the approximation is reasonable. This result shows that the
quintessence field value is increasing slowly as the age of the universe
increases. Then we obtain that the evolution of the quintessence energy
density $\rho _q$ as the cosmic scale factor $R$ is%
$$
\rho _q=\rho _{q0}\{1+\frac{6\tau }{\beta +4}\ln \frac R{R_0}\}^{-\beta /2},%
\eqno{(2)} 
$$
and the cosmic scale factor expands as the universe time%
$$
R=R_0\exp \{\frac{\beta +4}{6\tau }[(\frac t{t_0})^{4/(\beta +4)}-1]\},%
\eqno{(3)} 
$$
where the constant $\tau ^{-1}=\frac 32H_0t_0\simeq 1$ due to the matter
dominated$^{[9]}$, where $H_{0\text{ }}$is the Hubble constant in the time
of the beginning of the quintessence dominated and $t_0$ is the universe age
in the same time. It is obviously that the decreasing of the quintessence
energy density as the increasing of the cosmic scale factor is very slow,
its limited behavior is similar with the cosmological constant. Eqs.(2-3)
determined the future destiny of our universe. We can obtain the
equation-of-state of the quintessence as a function in the cosmic scale
factor 
$$
w_q(R)\equiv -1-\frac{d\ln \rho }{3d\ln R}=-[(4+6\ln \frac R{R_0})^{-1}\beta
+1]^{-1}.\eqno{(4)} 
$$
The equation-of-state of the quintessence in the beginning of the
quintessence dominated is obtained as $w_q^{(q)}=w_q(R_0)=-(\beta /4+1)^{-1}$%
.

The Type Ia Supernovas measured by Perlmutter et al. have average redshift $%
z\simeq 0.4$, in that epoch the universe undergoes the phase transition from
the matter dominated to the quintessence dominated. Watching on the
variation from the equation-of-state of the quintessence in the matter
dominated epoch $w_q^{(m)}$ to one in the quintessence dominated epoch $%
w_q^{(q)}$ carefully, we see that in the middle time, in which the matter
density is nearly equal to the quintessence density, the equation-of-state
of the quintessence should be $w_q^{(e)}=-(\beta /\gamma +1)^{-1}$, here we
take $\gamma =3$ may be a reasonable approximation. It is this
equation-of-state of the quintessence $w_q^{(e)}$ that should be applied in
the formula of the deceleration parameter of the recent observation of the
high redshift supernovas.

\section{Deceleration parameter}

{\hspace*{5mm}}Now we see the deceleration parameter%
$$
q_0=\frac 12\sum \Omega _i+\frac 32\sum w_i\Omega _i,\eqno{(5)} 
$$
where $\Omega _i\equiv \rho _i/\rho _c$ are the ratios of various component
densities to the critical density of the universe. We tested the correctness
of this formula, even if in the case of the existence of the quintessence.
We cite this correct formula directly, which appeared in Ref.[10].
Considering the red shift effect, we have%
$$
q_0=\frac 12\Omega _{m0}\cdot (1+z)^3+(\frac 12-\frac 32(\beta /\gamma
+1)^{-1})\Omega _{q0}\cdot (1+z)^{3-3/(\beta /\gamma +1)}.\eqno{(6)} 
$$
Noted the curvature term $\Omega _u$ with $w_u=-1/3$ has just be canceled.
When $\beta =0$ and $z=0.4$ we obtain $q_0\simeq \frac 43\Omega _{m0}-\Omega
_{\lambda 0}$, and Ref.[1] gives the observation result $0.6q_0=0.8\Omega
_{m0}-0.6\Omega _{\lambda 0}$ $=-0.20\pm 0.10$. In the later we shall use an
equivalent result eq.(1). Using the eqs.(6) and (1) one can discuss what
values the density ratios should take in the figure of $\Omega _{m0}$ and $%
\Omega _{\lambda 0}$, like the treatment in Ref.[1].

However we can think that as an approximation we can take $\Omega _{m0}=0.3$%
, $\Omega _{q0}=0.7$ at first, then use $q_0$ formula to constrain the
parameter $\beta $. Thus we obtain an important restriction, i.e., parameter 
$\beta $ is less than $2$. Here we used the assumption of the flat universe
predicted by the inflation models$^{[11]}$. Let us look at some concrete
data. For example, if $\beta =1$, then $w_q^{(e)}=-0.75$ and $q_0=-0.15$
which is out off the lower limit of the observation value; if $\beta =1/2$,
then $w_q^{(e)}=-0.86$ and $q_0=-0.22$ in according to eq.(6). If we take $%
\Omega _{m0}=0.2$, $\Omega _{q0}=0.8$, $z=0.4$, $\beta =2$, then $%
w_q^{(e)}=-0.60$ and $q_0=-0.20$; or we take $\beta =3$ and the same other
parameters, then $w_q^{(e)}=-0.50$ and $q_0=-0.06$. Anyhow the inverse power
index must be less than $3$. In fact it seems impossible that the sum of the
densities of the cold dark matter and the baryon matter is so small like $%
\Omega _{m0}=0.2$ in according to the estimated matter quantity from the
X-ray observation on the cluster of the galaxies$^{[12]}$, $\Omega
_{m0}=0.35\pm 0.07$.

If the inverse power index of the quintessence is $\beta =2$, the
equation-of-state of the quintessence in the radiation dominated era is $%
w_q^{(r)}=-1/3$, and the quintessence energy density is $\rho _q\propto
a^{-2}$, it decreases too slowly. At the early stage of the universe, due to 
$\rho _q/\rho _\gamma =(1+z)^{-2}$ and very high red-shift, require that the
quintessence energy density must be very low. In this case it is very easy
for the quintessence to produce overshot behavior and can not begin to track
even in very late time, therefore the initial condition can not be adjusted
in wide rang, and then it can not solve the coincidences problem. In fact
the ref.[4] has given a conclusion that the inverse power index must be
larger than $5$. Thus the quintessence with the low inverse power index lost
its attracting property.

During the derivation, we use some reasonable approximations, we think the
further exact results will not affect our main conclusion.

\section{Two term potential}

{\hspace*{5mm}}Of course we can use more complicated potential to overcome
this difficulty, for example $V=m_6^{10}\phi ^{-6}+m_2^6\phi ^{-2}$. When
field $\phi $ is small, the first term is dominated, it has a good tracking
behavior and wide adjusting rang of the initial condition. When field $\phi $
becomes large, the second term is dominated, the quintessence has the
suitable the equation-of-state $w_q$ for the deceleration parameter of the
supernovas. However, since the second term has a lower inverse power index,
the energy scale parameter $m_2$ has to be rather small, this may induce the
fine-turning problem. On the other hand, in this scheme, we must turn finely
the relative ratio of the mass parameters $m_6$ and $m_2$, let it undergo a
transition from the high inverse power term dominated to the low inverse
power term dominated just before the time of the matter-quintessence
equality.

Let us consider at what redshift this transition should happen. The tracking
solution should satisfy the equation$^{[4]}$%
$$
V^{\prime \prime }=\frac 92\frac{\beta +1}\beta (1-w_q^2)H^2,\eqno{(7)} 
$$
then we obtain the relation between the quintessential field value $\phi $
and the quintessential energy faction $\Omega _q$ in the matter-dominated
epoch 
$$
\phi ^2=\frac 23\beta (\beta +2)^2(\beta +4)^{-1}\Omega _{q0}(1+z)^{-2}M_p^2,%
\eqno{(8)} 
$$
When $\beta =6$, then $\phi _6\simeq 5\Omega _{q0}^{1/2}(1+z)^{-1}M_p$ and
when $\beta =2$, then $\phi _2\simeq 2\Omega _{q0}^{1/2}M_p$, the transition
should happen in the time of $\phi _6<0.6\phi _2$. Therefore we obtain the
redshift $z>3$. The more early this transition from high to low inverse
power terms process, the more narrow the adjusting rang of the initial
condition is. Why does the nature arrange these two transitions in such
order? 

The order magnitude of the mass parameters are $m_6\simeq 10^5$GeV and $%
m_2\simeq 10$MeV, if the low inverse power term is $m_1^5\phi ^{-1}$, then $%
m_1\simeq 1$keV. As a comparison, the cosmological constant $\Lambda =$$m_0^4
$ has $m_0\simeq 10^{-3}$eV. We see that the fine turning problem is relaxed$%
^{[13]}$.

\section{Conclusion}

{\hspace*{5mm}The important problem is whether the observation data can
constrain some models. We obtain the various equation-of-states of the
quintessence in the different stages of the universe, specially the
equation-of-states of the quintessence }$w_q^{(e)}=-(\beta /3+1)^{-1}$ {in
the time of the matter-quintessence equality. Using it we give the
restriction on the inverse power index $\beta \leq 2$} of the quintessence
potential from the deceleration parameter of the high redshift supernovas.

{The two functions of the tracking and the present quintessential energy
have to be achieved by the separated different two terms in the quintessence
potential respectively. Why the God takes so refinement arrangement?}

It is hopeful that the deceleration parameter which will be more accurate in
future, combining other observation, specially the total density ratio $%
\Omega _0=\Omega _m+\Omega _q$ from the position of the first Doppler peak
of CMBR$^{[14,15]}$, will give the more strict constrain on the potential
parameter of the quintessence.

If somebody does not approve of the non-terseness of the two term potential
of the quintessence, one can explore the exponential potentials or other
more complicated ones. To search new ideas to replace one of the
cosmological constant is the interesting challenge problem.

\bigskip

{\bf Acknowledgment:}

{\hspace*{5mm} This work is supported by The foundation of National Nature
Science of China, No.19675038 and No.19777103. The author would like to
thank useful discussions with Profs. J.R.Bond, L.Kofman, U.-L.Pen,
X.-M.Zhang Y.Z.-Zhang and X.-H.Meng. }

\bigskip

{\bf References:}

[1] S.J.Perlmutter et al., Nature 391(1998)51.

\qquad S.J.Perlmutter, et al., astro-ph/9812133.

[2] A.G.Riess, et al., Astron.J.116(1998)1009.

[3] B.Ratra and P.L.E.Peebles, Phys.Rev.D37(1988)3406.

[4] I.Zlatev, L.Wang and P.J.Steinhardt, Phys.Rev.Lett.896 (1999);

\qquad P.J.Steinhardt, L.Wang, and I.Zlatev, Phys.Rev.D59(1999)123504.

[5] M.S.Turner and M.White, Phys.Rev.D56(1997)4439.

[6] A.R.Liddle and R.J.Scherrer, Phys.Rev.D59(1999)23509.

[7] R.R.Caldwell, R.Dave and P.J.Steinhardt,

\qquad Phys.Rev.Lett.80(1998)1582.

[8] P.J.E.Peebles and A.Vilenkin, astro-ph/9810509.

[9] E.W. Kolb and M.S. Turner,

\qquad {\sl The Early Universe}, Addison Wesley, 1990.

[10] M.S.Turner, astro-ph/9904049; astro-ph/9912211.

[11] A.Linde, {\sl Particle Physics and Inflationary Cosmology},

\qquad 1990 by Harwood Academic Publishers.

[12] J. Mohr et al, Astrophys.J., in press (1999) (astro-ph/9901281).

[13] S.Weinberg, Rev.Mod.Phys.61(1989)61.

[14] M.Kamionkowski and A.Kosowsky, astro-ph/9904108.

[15] S. Dodelson and L. Knox, astro-ph/9909454.

\end{document}